\documentclass[
   10pt,%11pt, %12pt,
   twocolumn,
   showpacs,preprintnumbers,
   amsmath,amssymb,
   aps,
   prd,
]{revtex4-1}

\makeindex %it tells LaTeX to create the files needed for indexing
\usepackage{graphicx}% Include figure files
\usepackage[colorlinks=true, urlcolor=blue, linkcolor=blue, citecolor=blue]{hyperref}% add hypertext capabilities
\usepackage{definition}% Own definition
\begin{document}

\title{ Model of the AdS/QFT duality}
\author{ Stanis{\l}aw D. G{\l}azek and
Arkadiusz P. Trawi\'nski }
\affiliation{ Institute of Theoretical Physics,
          Faculty of Physics, 
          University of Warsaw,
          Ho\.za 69,
          00-681 Warsaw, Poland }
% It is always \today, today, but any 
%date may be explicitly specified
\date{ published 26 November 2013, \href{http://dx.doi.org/10.1103/PhysRevD.88.105025}{Phys. Rev. D 88, 105025 (2013)} }
%\doi{10.1103/PhysRevD.88.105025}

% PACS, the Physics and Astronomy Classification Scheme.
\preprint{IFT/13/04}
%\keywords{}

\begin{abstract}

It is observed and illustrated in a greatly simplified example 
that the idea of AdS/QFT duality can be considered a special 
case of the Ehrenfest's correspondence principle between 
classical and quantum mechanics in the context of relativistic 
dynamics of fields and renormalization group procedure for 
effective particles.

\end{abstract}

\maketitle

%\section{ The Ehrenfest equation }
%\label{sec:Ehrenfest}
%\subsection{ Explanation of Eq.~(\ref{eq:hadron}) }
%\label{sec:exp}
%\subsection{ Eigenvalue problem for a hadron state }
%\label{sec:eh}
%\subsection{Explanation of Eq.~(\ref{eq:Ehrenfest})}
%\label{sec:expEhrenfest}
%\subsection{ Averaging denoted by $\E{\ }$ }
%\label{sec:Averging}
%\section{The hadron form factors}
%\label{sec:hff}
%\section{ Model of the light-front holography }
%\label{sec:model}
%\subsection{ Transverse position variables }
%\label{sec:Space}
%\subsection{ Symmetric wave functions }
%\label{sec:Symetric}
%\subsection{ Expectation values in a model state }
%\label{sec:ExpModel}
%\subsection{ The model Ehrenfest function }
%\label{sec:ModelEf}
%\section{ Conclusion }
%\label{sec:conclusion}

%%%%%%%%%%%%%%%%%%%%%%
\section{Introduction}
%%%%%%%%%%%%%%%%%%%%%%

The Maldacena conjecture of AdS/CFT
duality~\cite{Maldacena:1997re} has been referred
to by Polchinsky and
Strassler~\cite{Polchinski:2001tt,
Polchinski:2002jw} to propose that hadrons may
correspond to classical fields in a
five-dimensional space-time. The fields can be
approximated by a product of a four-dimensional
plane wave and a nontrivial function of the fifth
coordinate. The nontrivial function is a solution
of the simple differential equation that depends
on the plane-wave four-momentum squared and hence
is capable of determining squares of masses of
hadrons through an adjustment of boundary
conditions in the fifth dimension. Irrespective of
the AdS/CFT conjecture and its M/string-theory
motivation, the question arises of how a solution
to quantum field theory (QFT) could reduce to a
solution of a simple, one-dimensional differential
equation. We address this issue using light-front
(LF) holography discovered by Brodsky and
de T\'eramond~\cite{deTeramond:2005su,Brodsky:2007hb,
deTeramond:2008ht}.

Brodsky and de T{\'e}ramond observed that the hadron form
factor formula proposed by Polchinsky and
Strassler matches the formula for hadron form
factors which one obtains in QFT when the latter is
developed in the front form (FF) of Hamiltonian
dynamics. The unique utility of the FF as potentially 
helpful in understanding duality is a fairly new 
finding in comparison with the fact that the FF of 
Hamiltonian dynamics as an intriguing alternative 
to the commonly used instant form (IF) had been 
discovered a long time ago by Dirac~\cite{Dirac:1949cp}. 

In this paper we find that the time-honored Ehrenfest 
correspondence principle between quantum and classical 
mechanics~\cite{Ehrenfest} can be used in combination 
with the LF holography to explain the possibility and 
actually suggest the necessity of the existence of the reduction 
of the QFT description of observables, such as hadronic 
masses squared or form factors, to the formulas that 
resemble the formulas inspired by the duality conjecture.

Brodsky and de T{\'e}ramond argue that the FF valence 
Fock sector wave function for a meson depends on 
the transverse separation between the quark and 
antiquark precisely in the same way that the AdS 
hadron field depends on the fifth-dimension 
coordinate. The specific soft-wall alteration of 
the AdS metric~\cite{Karch:2006pv} is meant to 
correspond to a harmonic oscillator potential 
between the two valence quarks in the direction 
transverse to the $z$ axis in the infinite momentum 
frame (IMF).

The interpretation of LF holography that the Ehrenfest
correspondence principle provides does not distinguish 
the Fock-space valence wave function. Instead of 
focusing on the valence Fock sector, the Ehrenfest 
correspondence relies on averaging over constituents 
in all Fock sectors in QFT. As a result, solutions to 
the Ehrenfest equation are expected to describe the 
charge distribution in hadrons in a scale-independent 
way, i.e., in a way which does not rely on the Fock-space 
decomposition at any arbitrarily selected 
renormalization-group scale. 

Regarding the renormalization-group scale dependence, 
it is known that the renormalization of Hamiltonians 
faces severe complications in the FF of dynamics~\cite{
Glazek:1993rc,Wilson:1994fk}. Therefore, we assume 
in our line of reasoning that in order to define the 
Fock-space decomposition of hadron states one can use 
the renormalization group procedure for effective 
particles (RGPEP)~\cite{Glazek:2011vg}. The most 
succinct description of the RGPEP can be found 
in Appendix C of Ref.~\cite{Glazek:2013gba}.

The paper is organized in the following way. Our
Ehrenfest equation is introduced in QFT in
Sec.~\ref{sec:Ehrenfest}. Hadron form factors are
discussed in Sec.~\ref{sec:hff}. The general
considerations of Secs.~\ref{sec:Ehrenfest} and
\ref{sec:hff} are subsequently illustrated in
Sec.~\ref{sec:model} with an example that
specifies key details of the required reasoning,
ignoring spin and all other such quantum numbers
of all constituents, i.e., treating all of them 
as indistinguishable scalar bosons. 
Section~\ref{sec:conclusion} concludes the paper 
with a summary of the connection between the Ehrenfest 
picture and LF holography.

%%%%%%%%%%%%%%%%%%%%%%%%%%%%%%%%%%%%%%%%%%%%%
\section{ The Ehrenfest equation for masses }
\label{sec:Ehrenfest}
%%%%%%%%%%%%%%%%%%%%%%%%%%%%%%%%%%%%%%%%%%%%%

We assume that the FF of dynamics in a QFT such as 
QCD allows one to represent physical states such as 
hadrons in a suitably defined Fock space once a theory 
is renormalized using the RGPEP. The RGPEP framework 
is close in its rules to the similarity renormalization 
group procedure introduced in Ref.~\cite{Glazek:1993rc} 
and used in Ref.~\cite{Wilson:1994fk}, but it includes 
an advantage of the nonperturbative operator calculus 
recently described in Refs.\cite{Glazek:2011vg,Glazek:2013gba}
in terms of simple examples.

For brevity and simplicity, we ignore below all quantum 
numbers of a hadron except its momentum. All the other 
quantum numbers are irrelevant to the idea of the Ehrenfest 
correspondence that we focus on. For the same reason, it 
is also useful to omit all quantum numbers of hadron 
constituents except their momenta. Despite these simplifications, 
we do continue to employ the word hadron as signifying a 
well-defined quantum mechanical representation of a physical 
particle as an eigenstate of the relevant Hamiltonian. The 
FF representation of a hadron is constructed in terms of an 
{\it a priori} infinite but convergent (thanks to the RGPEP) 
collection of the Fock-space components with the varying 
number $n$ of effective constituents and wave functions 
$\psi^{(n)}_P$ that correspond to the renormalization-group 
scale $\lambda$ (see Sec.~\ref{sec:exp} below),
\begin{align}
\label{eq:hadron}
  & \ket{\text{Hadron}\!:\! P^+,P^\perp} \\
  & \quad =  \sum_{n\,=\,n_\text{min}}^\infty\int\![{\boldsymbol p^\perp}\!,{\boldsymbol x}] 
       \,  \psi^{(n)}_{P}({\boldsymbol p^\perp}\!,{\boldsymbol x};\lambda)
       \ket{n :{\boldsymbol p^\perp}\!,{\boldsymbol x} P^+;\lambda}
   \nn\, .
\end{align}
%\text{\scalebox{1.5}{$\sigma$}}
For example, in the proton, one would have a 
combination of $\lambda$-dependent components 
$\ket{uud}$, $\ket{uudg}$, $\ket{uudq \bar q}$,
$\ket{uudgg}$, $\dots$ Thus, in baryons in QCD
one has $n_\text{min}=3$. In nonexotic mesons, 
$n_\text{min}=2$.

We shall define below a quantity which is an expectation 
value that describes a hadron and takes the form of a 
function that we shall call the Ehrenfest function. Our 
Ehrenfest function describes the motion of an averaged 
active parton, for example, the quark that absorbs 
a photon in a photon-hadron interaction described 
by the hadron form factor, with respect to spectators, 
averaged over all the Fock components. The averaging 
that we describe below yields the Ehrenfest equation 
that does not depend on $\lambda$ and takes the 
form of an eigenvalue problem for just one function 
$\psi(\vec k\,)$,
\begin{align}
\label{eq:Ehrenfest}
 \left[
       { \v k^2 }
       +  (m_\text{active}+m_\text{core})^2          
       +  U_\text{eff}
       \right]
       \psi(\v k)
& =
 M^2 \psi(\v k)
\, .
\end{align}
The precise meaning of the three-dimensional variable 
$\v k$ will follow from our analysis of motion 
of the active constituent with respect to the rest 
of a hadron. The effective Ehrenfest potential 
$U_\text{eff}$ shall be discussed at length.

In order to outline the procedure of averaging over 
$\lambda$-dependent Fock components and over motion 
of all spectator partons in them, this article sketches 
the scenario in which three different equations appear: 
the Schr\"odinger-picture Hamiltonian eigenvalue equation 
in QFT in the FF of dynamics, the Ehrenfest equation 
that results from our averaging, and the form factor 
formula that Brodsky and de T\'eramond use in their 
holographic picture with a soft-wall, or a harmonic 
oscillator potential. As a result, we suggest that 
the classical five-dimensional equation with a potential 
such as in the soft-wall model~\cite{Karch:2006pv}
corresponds to our Ehrenfest equation that resembles
the Schr\"odinger equation for one particle in a
corresponding potential. 

%%%%%%%%%%%%%%%%%%%%%%%%%%%%%%%%%%%%%%%%%%%%%%%%%%%
\subsection{ Explanation of Eq.~(\ref{eq:hadron}) }
\label{sec:exp}
%%%%%%%%%%%%%%%%%%%%%%%%%%%%%%%%%%%%%%%%%%%%%%%%%%%

The hadron state can be written in terms
of the FF wave functions that depend 
on the transverse momenta of the constituents,
${\boldsymbol p^\perp} = (p_i^\perp)_{i=1,\dots,n}$,
and ratios of longitudinal momenta of 
constituents to the hadron longitudinal 
momentum, ${\boldsymbol x} = (x_i)_{i=1,\dots,n}$, 
$x_i = p_i^+/P^+$. In our notation, bold symbols 
will always refer to a set of variables that 
describe a set of constituents. For every
constituent, its FF fraction $x$ matches its
parton-model Bjorken variable $x$ in the IMF.

Let us ignore all details and assume that all 
constituents have the same mass $m = m(\lambda)$ 
so that one can define their free minus momentum 
components using $p_i^2 = p_i^+p_i^- - p_i^{\perp\,2} 
= m^2$. For integrating over kinematical momentum 
variables, we use notation
\begin{align}
\int\![\boldsymbol p^\perp]
   & = \prod_{i=1}^n \int\!{d^2 p_i^\perp \over (2\pi)^2}
   & \text{and} &&
\int\![\boldsymbol x]
   & = \prod_{i=1}^n \int\!{d p_i^+ \over (2\pi)2 p_i^+} \,.
\end{align}
In order to model QFT such as QCD, we assume that 
for $\lambda \sim \Lambda_{QFT}$, where $\Lambda_{QFT}$ 
stands for a characteristic physical parameter of a 
theory (such as $\Lambda_{QCD}$ in a specific 
renormalization scheme, which in our case is the 
RGPEP), a model of a hadron can be constructed 
using a small number of constituents. 
For example, in a proton the sector with $n = 3$ and 
$\lambda \sim \Lambda_{QCD}$ almost saturates the 
entire sum over $n$, gluons contributing to the bulks 
of constituent quarks, e.g., see Ref.~\cite{Glazek:2011vg}. 
In great contrast, the expansion into Fock components 
corresponding to $\lambda \gg \Lambda_{QFT}$ extends 
over a whole range of $n$-particle states with large 
numbers $n$ and large relative momenta.

Our model of a hadron state is normalized using the condition
\begin{align}
   \label{eq:normalization}
   &\braket{\text{Hadron}\!:\!P'^+,P'^\perp}{\text{Hadron}\!:\!P^+,P^\perp}\\
   & \quad = 2P^+ (2\pi)^3 \delta(P'^+ - P^+)\, \delta^{(2)}(P'^\perp - P^\perp)
   \nn\, .
\end{align}
The relative momenta are separated from the
total momentum by writing
\begin{align}
\label{eq:psin}
   &  \psi^{(n)}_{P}({\boldsymbol p^\perp}\!,{\boldsymbol x};\lambda)
      = 2(2\pi)^3\delta\Big(\sum_{i=1}^n x_i - 1
     \Big)\delta^{(2)}\Big(P^\perp - \sum_{i=1}^n p_i^\perp\Big) \nn\\
   & \hspace{7em} \times
            \psi^{(n)}({\boldsymbol k^\perp}\!,{\boldsymbol x},;\lambda) \,,
\end{align}
where ${\boldsymbol k^\perp} = (k_i^\perp)_{i=1,\dots,n}$
and $k_i ^\perp  = p_i^\perp - x_i P^\perp$ is the 
transverse relative momentum of the $i$th constituent 
with respect to the \emph{center of mass of
constituents}
treated as free particles of mass $m$.
The wave function $\psi^{(n)}
({\boldsymbol k^\perp}\!,{\boldsymbol x},;\lambda)$
always appears in conjunction with the $\delta$ functions 
such as in $\psi^{(n)}_{P}({\boldsymbol p^\perp}\!,{\boldsymbol x};
\lambda)$ in Eq.~(\ref{eq:psin}). Therefore,
its domain is actually limited by the conditions
\begin{align}
\label{eq:domain}
   \sum_{i=1}^n x_i
   & = 1 
   & \text{and} &&
   \sum_{i=1}^n k_i^\perp
   & = 0 \, .
\end{align}
Thus, only $n-1$ of the $n$ three-dimensional 
arguments of $\psi^{(n)} ({\boldsymbol k^\perp}\!,
{\boldsymbol x},;\lambda)$ are independent. 

In every sector, we distinguish one constituent 
and we focus on the description of its motion with
respect to other constituents. The selected 
constituent is described using variables with 
subscript $n$ or without any subscript. The other 
constituents are named spectators or a core, 
depending on the context, and they are described
using variables with subscripts $i = 1$, $\dots$, $n-1$. 

%%%%%%%%%%%%%%%%%%%%%%%%%%%%%%%%%%%%%%%%%%%%%%%%%%%%
\subsection{ Eigenvalue problem for a hadron state }
\label{sec:eh}
%%%%%%%%%%%%%%%%%%%%%%%%%%%%%%%%%%%%%%%%%%%%%%%%%%%%

In QFT, the FF Hamiltonian $\hat P^-$ determines 
the structure of a composite system, as a solution 
to its eigenvalue equation. The eigenvalue is 
expressible in terms of the components $P^+$ and 
$P^\perp$ of the total kinematical momentum of 
the system and its mass squared, $M^2$. We call 
such a solution a hadron when the eigenvalue $M^2$ 
takes one of the several smallest possible values 
for a system with the hadron quantum numbers. The 
limitation to smallest masses is meant to exclude 
from consideration the scattering and bound states 
of entire hadrons. So,
\begin{align}
   & \hat P^- \ket{\text{Hadron}\!:\!P^+,P^\perp } \\
   & \quad = {M^2 + P^{\perp \, 2} \over P^+}
       \ket{\text{Hadron}\!:\!P^+,P^\perp } \nn\,.
\end{align}
$\hat P^-$ is a sum of a free part $\hat P_0^-$
ascribing free FF energies to the system constituents 
and the interaction part $\hat P_I^-$. Therefore, we 
can write the expectation value of $P^+ \hat P^- -  
P^{\perp \, 2}$ in a hadron in the form
\begin{align}
   \label{eq:Schordginger}
   M^2 
   & = 
       \,\sum_{n}
       \int\![\boldsymbol\kappa^\perp\!,{\boldsymbol \chi}]  
       \,\,2(2\pi)^3\delta\Big(\sum_{j=1}^{n-1} \chi_j -1\Big)\delta^{(2)}
\Big(\sum_{j=1}^{n-1} \kappa_j^\perp\Big)
   \nn \\ & \quad \times
   \int_{k,x} \psi^{\dagger(n)}_{k,x}
   \Big[{k^{\perp\,2}\over x(1-x)} 
         + {m_n^2 \over x}
         + {\cM_{n-1}^2 \over 1 -x}
   \Big]
   \psi^{(n)}_{k,x} \nn\\
    & \quad + (\text{connected interactions})  \, .
\end{align}
The integration symbol $\int_{k,x}$ appears
here and later on universally in the entire 
paper to denote the integration over subscript
arguments $k^\perp$ and $x$ in the Fock-space 
wave functions of a hadron state, $\psi_{k,x}^{(n)}$, 
as required, and over arguments of our Ehrenfest 
function, see Eq.~(\ref{eq:averaging_free}). In 
all the cases,
\begin{align}
\int_{k,x}
& =
\int\!{d^2k^\perp \, dx \over 2(2\pi)^3 x(1-x)} \, .
\end{align}

For every $n$, the set of variables denoted by 
symbols $\boldsymbol{k^\perp}$ and $\boldsymbol{x}$ 
defined previously for $n$ constituents is split 
in Eq.~(\ref{eq:Schordginger}) into $k^\perp = k_n^\perp$ 
and $x = x_n$ for the active constituent and a 
set of variables denoted by $\boldsymbol{\kappa^\perp} =
(\kappa^\perp_i)_{i,\dots,{n-1}}$ and $\boldsymbol{\chi}
= (\chi_i)_{i=1,\dots,n-1}$ for the $n-1$ spectators
that form the core. The constituents' relative momentum 
variables internal to the core are defined by 
$\kappa^\perp_j = k^\perp_j + \chi_jk^\perp$.

Note the plus sign in front of the terms $\chi_j k^\perp$
which correspond to $x_iP^\perp$ in the definition of 
$k_i^\perp$ below Eq.~(\ref{eq:psin}). The change in
sign results from the core being made of spectators 
carrying as a whole $-k^\perp$ and $1-x$ when the active 
constituent carries $k^\perp$ and $x$. 

The functions $\psi^{(n)}_{k,x}$ depend on $x$ and $k^\perp$ 
in a way that we separate in our notation from their
dependence on the variables $\boldsymbol \kappa^\perp$ 
and $\boldsymbol{\chi}$. The function
\begin{align}
   \label{eq:m_n-1}
   \cM_{n-1}^2 (\boldsymbol\kappa^\perp\!,{\boldsymbol \chi})
   & = \sum_{j=1}^{n-1} {\kappa_j^{\perp \, 2} + m_n^2\over \chi_j } 
\end{align}
is an invariant mass squared of the $n-1$ 
spectators that form a core in the sector 
number $n$, treated as a set of free particles 
of mass $m_n$. 

Both the value of $m_n^2$ and the presence of connected 
interactions in Eq.~(\ref{eq:Schordginger}) require an 
explanation. The masses result from adding the self-interactions 
generated in the eigenvalue equation to the renormalized mass 
squared, $m^2$, that one obtains in $\hat P^-$ at scale $\lambda$ 
using the RGPEP. The connected interactions are the ones that 
are left after the constituent self-interactions are included 
in their mass terms.

Consider first a simple theory in 
which the Hamiltonian contains $m^2$ and its 
eigenstate is built from only a two- and a 
three-constituent sector, so that
\begin{align}
\ket{\psi} & = \ket{2} + \ket{3} \, .
\end{align}
The Ehrenfest expectation value of the Hamiltonian 
$ \hat P^- = \hat P_0^- + \hat P_I^-$ has the form
\begin{align}
\label{eq:evH}
\bra{\psi} \hat P^- \ket{\psi} 
& = \bra{3} \hat P^-  \ket{3} + 
    \bra{2} \hat P^-  \ket{2} \\
& + \,\bra{3}\hat P^-_I\ket{2} + \bra{2}\hat P^-_I\ket{3} \nn \, .
\end{align}
In the matrix elements that are diagonal in the number
of constituents, the interaction terms change the momenta
of at least two constituents on at least one side of the 
matrix element, because $m^2$ in $\hat P^-_0$ includes 
all terms that contribute to single-particle energies 
in $\hat P^-$. The matrix elements that involve a change 
of the number of constituents can be expressed in terms 
of diagonal matrix elements using the eigenvalue equation 
for the state $\ket{\psi}$. In the simple model, one can 
assume that the interactions in the component $\ket{3}$ 
are negligible in comparison to relevant eigenvalues 
of $\hat P_0^-$ and write
\begin{align}
\ket{3} & \simeq {P_3 \over P^- - \hat P^-_0} \hat P^-_I\ket{2} \, ,
\end{align}
where $P_3$ denotes projection on the three-body component. 
The presence of an eigenvalue in the denominator can be nearly 
ignored for a whole set of lightest hadron masses $M$ while 
the total momentum eigenvalues $P^+$ and $P^\perp$ drop out 
irrespective of the value of $M$. Assuming the above 
approximations, the expectation value of Eq.~(\ref{eq:evH}) 
takes the form 
\begin{align}
\label{eq:evH1}
\bra{\psi} \hat P^- \ket{\psi} 
& \simeq \bra{3} \hat P^-_0 \ket{3} + \bra{2}\hat P^-\ket{2} \\
& \quad +\ 2 \ \bra{2}\hat P^-_I 
                      { P_3 \over P^- - \hat P^-_0} 
                      \hat P^-_I \ket{2} \nn \, .
\end{align}
The last term contains interaction effects due to the
exchange of field quanta between two constituents and 
the self-interaction terms that change $m^2$ to 
$m_2^2$ in the sector $\ket{2}$ and make $m_2^2$  
differ from $m_3^2 = m^2$ in the sector $\ket{3}$. 

Now, in the case of a complex theory, the number 
of effective-particle sectors and the number of 
interaction terms in a $\hat P^-$ corresponding to any 
finite value of $\lambda$ will both be in principle 
infinite, although finite $\lambda$ typically implies 
a quick decrease of wave functions with an increase of 
the number of massive constituents and their invariant 
mass (this is a general feature of Hamiltonians
one generates using the RGPEP). In any case, $\hat P^-$ 
in theories of physical interest contains the terms 
that change a single constituent to more than one of 
them or turn more than one constituent into 
just one. All these terms will result in sector-dependent
single-particle mass-squared terms, which 
combine with $m^2$ and are denoted in 
Eqs.~(\ref{eq:Schordginger}) and (\ref{eq:m_n-1}) 
by $m_n^2$. 

%%%%%%%%%%%%%%%%%%%%%%%%%%%%%%%%%%%%%%%%%%%%%%%%%%%%
\subsection{Explanation of Eq.~(\ref{eq:Ehrenfest})}
\label{sec:expEhrenfest}
%%%%%%%%%%%%%%%%%%%%%%%%%%%%%%%%%%%%%%%%%%%%%%%%%%%%

Equation~(\ref{eq:Schordginger}) indicates that QFT
quite generally yields an averaged equation of motion 
of an active constituent with respect to the core 
formed by spectators. We call this averaged equation 
the Ehrenfest equation, Eq.~(\ref{eq:Ehrenfest}), 
because of an analogy to the Ehrenfest equation 
for expectation values of observables in quantum
mechanics. The original Ehrenfest equation for 
expectation values in quantum mechanics corresponds 
to the Newton equations of classical mechanics~\cite{Ehrenfest}. 
In our case, the eigenvalue equation for a Hamiltonian 
in QFT implies an expectation-value equation which 
results from averaging over all numbers, momenta, and 
other quantum numbers of virtual constituents. Our 
Ehrenfest equation thus explains in what sense a QFT 
can be dual to a classical theory and how it may 
render a holographic picture of hadrons.

Since the expectation value of Eq.~(\ref{eq:Schordginger}) 
involves averaging over the structure and dynamics inside 
the core and between the active constituent and the core, 
our Ehrenfest equation describes the motion of the active 
constituent in the effective potential that describes the 
net result of all connected interactions in the hadron. The 
effective potential is denoted below and in Eq.~(\ref{eq:Ehrenfest}) 
by $U_\text{eff}$. 

We first focus on the kinetic and mass terms that 
result from averaging in Eq.~(\ref{eq:Schordginger}).  
These terms help us define the variables that are 
suitable for writing our Ehrenfest equation in a 
simple form. Namely, the first term on the right-hand 
side of Eq.~(\ref{eq:Schordginger}) is written as
%\begin{widetext}
\begin{align}
   \label{eq:averaging_free}
   & \E{ \int_{k,x} \psi^{\dagger(n)}_{k,x}
         \left[ {k^{\perp\,2} \over x(1-x)} + {m_n^2 \over x} + {\cM_{n-1}^2
             \over 1- x} \right]
         \psi^{(n)}_{k,x}
     }\\
   & \hspace{-1em} =  \int_{k,x}
       \psi(k^\perp\!,x)^\dagger
        \left[ {k^{\perp\,2} \over x(1-x)} + {m_\text{active}^2 \over x} 
          + {m_\text{core}^2 \over 1- x} \right]
       \psi(k^\perp\!,x) \, . \nn   
\end{align}
%\end{widetext}
The brackets $\E{~}$ denote integrating over 
the relative motion of constituents and summing 
over all sectors that form a hadron. We call the 
function $\psi(k^\perp\!,x)$ the Ehrenfest function.
It depends on three momentum variables, two transverse,
$k^\perp = (k^1, k^2)$, and one $+\,$ fraction, $x$, 
no matter how many constituents of any kind a true 
hadron contains in any of its Fock components in 
any QFT.

The same averaging produces all the quantities that 
appear in our Ehrenfest equation. In the model example 
of Sec.~\ref{sec:model}, the averaging will be defined 
in detail using Eq.~(\ref{eq:EX}). Here we begin our
description of the averaging by stating that it satisfies 
the following conditions:
\begin{subequations}
   \label{eq:averaging}
\begin{align}
\E{ \int_{k,x} \psi^{\dagger(n)}_{k,x}\ \varkappa_n^2\ \psi^{(n)}_{k,x}}
   & = \varkappa^2 \,,
   \\
\E{ \int_{k,x} \psi^{\dagger(n)}_{k,x}\ m_n^2\ \psi^{(n)}_{k,x}}
   & = m_\text{active}^2 \,,
   \\
\E{ \int_{k,x} \psi^{\dagger(n)}_{k,x}\ \cM_{n-1}^2\ \psi^{(n)}_{k,x}}
   & = m_\text{core}^2\, .
\end{align}
\end{subequations}
The width $\varkappa$ in 
Eq.~(\ref{eq:averaging}a) corresponds to the 
half-width of the Ehrenfest function, whereas 
the quantities $\varkappa_n$ correspond to the 
half-widths of the Fock-space wave functions 
$\psi_{k, x}^{(n)}$. Below, we show in 
Eq.~(\ref{eq:FormFactorE}) that $\varkappa$ is 
an observable corresponding to the inverse of 
a hadron size. The hadron size depends on all 
the Fock sectors, whose wave functions depend 
on $\lambda$, but the hadron size is an observable 
measurable in terms of form factors and it does 
not depend on $\lambda$. Hence, $\varkappa$ must 
not depend on our RGPEP scale $\lambda$. The mass 
terms in Eqs.~(\ref{eq:averaging}b) and~(\ref{eq:averaging}c)  will be 
discussed later.

In addition to the kinetic and mass terms, we
have to include the expectation value of all 
the connected interactions that yield the effective 
interaction potential between the active constituent 
and the core. We call this potential the Ehrenfest 
effective potential, or just the Ehrenfest potential. 
We write
\begin{align}
   \label{eq:Ueff}
   U_\text{eff}
   &= \Big\langle\text{connected interactions}\Big\rangle \, .
\end{align}
This completes our introductory description of
how we evaluate the relevant expectation values.

Variation of the Ehrenfest expectation value 
with respect to $\psi(k^\perp\!,x)$, keeping 
the norm of $\psi(k^\perp\!,x)$ fixed, yields 
Eq.~(\ref{eq:Ehrenfest}). We now proceed to the 
description of our averaging procedure in greater 
detail.

%%%%%%%%%%%%%%%%%%%%%%%%%%%%%%%%%%%%%%%%%%%%%%%%%%%%%%
\subsection{ Averaging denoted by $\langle~\rangle$ }
\label{sec:Averging}
%%%%%%%%%%%%%%%%%%%%%%%%%%%%%%%%%%%%%%%%%%%%%%%%%%%%%%

First of all, we see no reason for the Ehrenfest 
function $\psi(k^\perp\!,x)$ to depend on the 
RGPEP scale $\lambda$, since its modulus squared 
turns out to describe the measurable charge 
distribution in a hadron (e.g., see Sec.~\ref{sec:hff}
below). As so closely related to observables, the 
Ehrenfest function may only involve some width 
parameter $\varkappa$ that characterizes the QFT,
in which conformal symmetry is naturally broken. 

For example, a natural breaking of conformal symmetry 
occurs in the case of asymptotic freedom, as in QCD, 
where $\varkappa$ must be in a one-to-one correspondence 
to $\Lambda_{QCD}$. Since the canonical QCD Lagrangian 
density does not contain $\Lambda_{QCD}$, one needs to 
consider renormalization in order to see its presence
(in the FF of QFT, one may introduce $\Lambda_{QCD}$ 
using the RGPEP). We do not consider renormalization in 
its full generality below but we do address the issue of 
dependence of effective theories on the RGPEP scale 
$\lambda$. It is in this dependence where one expects 
the parameters such as $\varkappa$ or $\Lambda_{QCD}$ 
to appear as reference quantities. These reference 
parameters allow us to tell the magnitude of $\lambda$ 
and consider nontrivial functions of $\lambda$.

The averaging over Fock sectors implies that 
the Ehrenfest function is not equal to the 
Fock-space wave function in the sector with a 
smallest admissible number of constituents in 
a hadron, which we denote by $\psi_{k, x}^{(n)}$ 
with $n = n_{\text{min}}$. The function $\psi_{k, x}
^{(n_{\text{min}})}$ changes with $\lambda$ 
considerably. In contrast, our Ehrenfest 
function is independent of $\lambda$. 

Regarding the issue of connection with local QFT,
we wish to stress that our Ehrenfest function
$\psi(k^\perp\!,x)$ is not identifiable with
$\psi_{k, x}^{(n_{\text{min}})}$ in the canonical
theory that is regularized and supplied with a
complete set of counterterms. Such theory only
forms the initial condition for the RGPEP
evolution of effective theory with the scale
parameter $\lambda$, at the initial value of
$\lambda = \infty$. In such initial-condition
theory, the canonical quanta appear in hadrons in
great numbers and with great relative momenta. The
canonical quanta could also be called the current
quanta, in analogy with quarks in the formal
current algebra. The canonical quanta could also
be thought about as quarks and gluons considered
in the perturbative QCD and parton models. 

In contrast, our Ehrenfest function is a
relatively soft function of $k^\perp$ and $x$ over
a dominant range of these variables. It satisfies
a simple wave equation with a soft potential.
Thus, the Ehrenfest function for a hadron can only
be seen as, graphically speaking, the function
whose modulus squared provides an averaged parton
charge distribution. It does not report on any
details of potentially violent interactions of
quanta for any large $\lambda$, i.e., much larger
than the parameters such as $\varkappa$,
$\Lambda_{QCD}$, or quark masses in the
constituent quark model. 

On the other hand, we wish to clarify that when 
$\lambda$ is comparable with quark masses in the 
constituent quark model, the effective-particle 
sector with $n = n_{\text{min}}$ 
may saturate to a large extent the probability 
distribution of constituents in a hadron
in the effective-particle Fock-space basis.
In this case, our Ehrenfest function in the
nonrelativistic domain of $k^\perp$ and $x$ may 
be related to $\psi_{k, x}^{(n_{\text{min}})}$ in the 
relatively simple way that involves mainly 
averaging over motion and numbers of a small 
set of effective constituent components in a 
hadron. Thus, one can expect a close resemblance 
between our Ehrenfest function and the quantum
mechanical wave functions in constituent 
quark models in the domain of nonrelativistic 
relative motion of an effective quark with
respect to other effective constituents, when
all these effective particles correspond
to a small value of $\lambda$. 

We also wish to stress that despite the simplicity 
of our Ehrenfest function, it is compatible with 
the hadron form factors being dependent on all Fock 
components. We shall see below that the Ehrenfest 
function includes contributions from all sectors and 
sums them up in the form factor formulas as required 
by the renormalized theory with an arbitrary value 
of $\lambda$, for as long as the current operators 
evolve with $\lambda$ without involvement of significant 
form factors for the constituents themselves. 

The Ehrenfest interpretation of holography suggested 
in this paper is thus alternative to the one that is 
based on the valence Fock-space wave function, i.e., 
actually, the wave function with $n = n_{\text{min}} = 2$ 
in mesons. The valence interpretation was suggested in
Refs.~\cite{deTeramond:2013it,Brodsky:2013npa} using 
Ref.~\cite{Pauli:1998tf}. The latter work attempts to 
reduce the hadron eigenvalue problem to some equation 
for the valence Fock wave function in the canonical QFT. 
For this purpose, Ref.~\cite{Pauli:1998tf} employs 
Eq.~(2.7) of Ref.~\cite{Krautgartner:1991xz}. But Eq.~(2.7)
of Ref.~\cite{Krautgartner:1991xz} is used there to 
replace the eigenvalue in a hadron eigenvalue problem 
by a combination of free energies of arbitrarily moving 
constituents, when one eliminates Fock sectors trying to 
obtain an equation for the valence sector alone. Since 
the canonical parton energies in QFT such as QCD can 
differ from the hadron eigenvalue by arbitrarily large 
amounts, it is not clear how the hadron eigenvalue problem 
could be reduced to its valence component alone in QFT 
using the approach proposed in Refs.~\cite{Pauli:1998tf,
Krautgartner:1991xz}. 

Considering the alternative interpretation of LF holography 
proposed in this paper, one also needs to take into account 
that the valence component is generally not sufficient to 
evaluate form factors; all Fock sectors contribute. This 
means that the valence, or just twe two-parton component of a
hadron cannot appear alone in the exact representation of 
hadron form factors that is claimed valid on the basis of 
duality arguments~\cite{Polchinski:2002jw,Polchinski:2001tt}. 
Therefore, our Ehrenfest interpretation of holography 
and duality through averaging over parton numbers and 
motion of spectators within the RGPEP appears to resolve 
the conceptual difficulty with the identification of  
$\psi(k^\perp\!,x)$ with the Fock-space wave function 
for $n = n_{\text{min}}$ in QFT. 

In its most simple version at small $\lambda$, the need 
for our alternative interpretation is evident in the case 
of baryons, where $n_{\text{min}}=3$ and one has to perform 
averaging over spectators in the form factor formula in
order to obtain an expression that only involves a function 
of just three variables $k^\perp$ and $x$, instead of the 
six momentum variables that form the arguments of the valence 
wave function for constituent quarks in baryons in QCD with 
small $\lambda$ (see Ref.~\cite{Glazek:2011vg} for the 
relevant discussion using the RGPEP). Note that the 
constituent picture at small $\lambda$ incorporates the 
entire Fock-space composition of a hadron in a canonical 
theory within the structure of the small-$\lambda$ constituent 
quarks.

Now we proceed to an explanation of the averaging that
renders mass parameters. The quantities $m_\text{active}^2$ 
and $m_\text{core}^2$ in Eqs.~(\ref{eq:averaging}b) and 
(\ref{eq:averaging}c) are the expectation values of the 
active constituent's and core's masses squared, respectively. 
For example, in QCD, for mesons built from lightest quarks, 
one may expect $m_\text{active} \sim m_\text{core} \sim 
\Lambda_{QCD}$. This is expected by analogy with the 
constituent quark model, in accordance with the idea 
that the constituent quark mass corresponds to $m$ at 
$\lambda \sim \Lambda_{QCD}$, at which $\lambda$ the 
effective quark self-interactions are small. By the same 
token, in the light baryons $m_\text{core}$ is expected 
to be about twice larger than $m_\text{active}$. For 
heavy quarks, variation of the mass $m$ with $\lambda$ 
is not as significant as for the light ones and the 
averaged masses are expected similar to the ones assumed 
in QCD.

Following~\cite{Glazek:2011vg}, we introduce the third 
component of relative momentum of the active constituent 
with respect to the core by writing
\begin{align}
   \label{eq:identity}
   {m_\text{active}^2 \over x} + {m_\text{core}^2 \over 1 - x}
   & = {\left[ m_\text{core} \, x - m_\text{active} \, (1-x) \right]^2 \over x(1-x)} \\
   & \quad + (m_\text{active} + m_\text{core})^2 \, , \nn
\end{align}
and identifying in the first term on the right-hand side
the square of 
\begin{align}
\label{eq:k3}
   k^3
   & = m_\text{core} \, x - m_\text{active} \, (1-x) \, .
\end{align}
This variable is introduced as an intermediate
step in defining $k_x$, $k_y$, and $k_z$ in the 
Ehrenfest equation by the formula
\begin{align} 
\label{eq:veck}
   \v k
   & = (k_x,k_y,k_z) = {(k^\perp\!,k^3)\over \sqrt{ x(1-x)}}\, .
\end{align}
Note that the free invariant mass squared of the
active constituent and core on the right-hand 
side of Eq.~(\ref{eq:averaging_free}) can be 
written in terms of $\v k$ as
\begin{align}
\label{eq:Jacobi}
& {k^{\perp\,2} \over x(1-x)} + {m_\text{active}^2 \over x} +
          {m_\text{core}^2 \over 1- x} \\ 
& \hspace{5em} = 
\v k^2 + (m_\text{active} + m_\text{core})^2 \nn\, .
\end{align}
The key feature of the variable
$\v k$~\cite{Glazek:2011vg} is that a simple rescaling 
relates it to the Jacobi relative momentum 
of an effective active constituent with respect 
to the corresponding effective core in the 
nonrelativistic domain of their relative 
motion around the minimum of their potential 
energy described by $ U_\text{eff}$ in 
Eq.~(\ref{eq:Ehrenfest}). The required 
rescaling factor is the root of the ratio
of the reduced mass $\mu =  m_\text{active}
\, m_\text{core}/(m_\text{active} + m_\text{core})$, 
to the sum of masses, $m_\text{active} + m_\text{core}$. 
Thus, the Jacobi relative momentum of constituents 
in nonrelativistic two-body models of hadrons 
is $\sqrt{\beta(1-\beta)}\, \v k$ where $\beta 
= m_{\text{active}}/(m_{\text{active}} + m_{\text{core}})$. 

We mention the rescaling because in interpreting 
our result for the relativistic Ehrenfest function, 
such as in Eqs.~(\ref{eq:oscillatorgs}), 
(\ref{eq:psis}), or (\ref{eq:psivk}) below, the 
factor $\sqrt{\beta(1-\beta)}$ must be taken 
into account in order to connect the relativistic 
FF theory with nonrelativistic models of hadrons 
for $|\v k|$ smaller than the masses $m_\text{active}$ 
and $m_\text{core}$. Physically, the rescaling is 
required due to the fact that the nonrelativistic 
expression for energy of a system of two free
objects at rest as a whole is $\cM = m_\text{active} 
+ m_\text{core} + \v k^2/(2\mu)$ and leads to $\cM^2 
= (m_\text{active} + m_\text{core})^2 + \v k^2/
[\beta(1-\beta)]$ with corrections order $\v k^4$ 
that are neglected in the nonrelativistic models. 
In contrast, the FF expression for $\cM^2$ given 
in Eq.~(\ref{eq:Jacobi}) is exact for arbitrary values
of $\v k$.

The sum of masses $m_\text{active} + m_\text{core}$
that appears squared in the second term in 
Eq.~(\ref{eq:identity}), differs from the mass 
eigenvalue, $M$. Besides the relative motion 
of the active constituent and core, the difference
results also from the potential energy $U_\text{eff}$ 
that is obtained from averaging of the interactions, 
Eq.~(\ref{eq:Ueff}). The averaging of the interactions 
is carried out excluding the disconnected self-interactions 
that we have already included in the masses $m_n$. 
The effective potential is further described 
using Eq.~(\ref{eq:harmonic_potencial}) below.

Regarding the shape of $U_\text{eff}$ as a function 
of the average distance between the active constituent
and the core, one can expect that in the ground 
states and lowest excited states of hadrons it is 
quadratic. The quadratic potential is expected
as a result of the standard reasoning in which one 
assumes that every constituent moves in a field 
produced by many others, such as in the case of a 
nucleon in a nucleus. The constituents interact with 
each other and every one of them sees the potential 
produced by others. One can think about the self-consistent 
potential that binds every constituent. This potential 
has a minimum, and around a physically reasonable minimum 
a potential is most likely quadratic. Precise mathematical 
reasoning would require modeling of a lot of details as 
in nuclear physics. Let us consider instead a simple 
model of a chargeless hadron in which in every Fock 
component all constituents interact through a Coulomb 
potential. One can imagine a small parton of charge $Q$ 
moving in a relatively large, uniformly charged sphere 
of spectators with total charge $-Q$ and density $\rho$. 
Then, the Coulomb potential the parton sees at a distance 
$r$ from the spectator-sphere center is the charge in the 
sphere of radius $r$, $Q_r = \rho$ times $4\pi r^3/3$, 
divided by $r$. Such potential is of the harmonic oscillator 
type. This picture could correspond to large $\lambda$
in the RGPEP so that the Fock components with many partons 
are dominant. At small $\lambda$, one may rather think in 
terms of two large constituents that overlap nearly 
entirely~\cite{Glazek:2011vg}. If these constituents are 
each described as Gaussian charge densities of opposite 
signs and with their centers separated by a distance 
much smaller than their individual sizes, the potential 
between them is also of the harmonic oscillator type as 
long as the opposite-sign charge densities attract each 
other via any reasonable potential, including the Coulomb 
one. Thus, at both ends of the RG trajectory in $\lambda$ 
the oscillator potential appears reasonably realistic
if the Coulomb potential is considered realistic. The 
lack of scale dependence of the final Ehrenfest oscillator 
must result from summing contributions from all sectors 
at every scale $\lambda$, leading to the same averaged
$U_\text{eff}$ no matter what value $\lambda$ has.

Irrespective of the details introduced in the
above examples, the bottom-line argument for the
harmonic oscillator potential between the active
constituent and the core in the ground states and
lowest excited states of hadrons is that the
potential should describe the constituent motion
around a minimum of their potential energy. This
principle appears to us sufficient for suggesting
the quadratic form for $U_\text{eff}$ because we
cannot identify any reason for the minimum to be
described by the fourth or higher even power of
the distance. Hence, $U_\text{eff}$ around its 
minimum should have the form
\begin{align}
   \label{eq:harmonic_potencial}
   U_\text{eff}
   & = -\varkappa^4\left({\partial \over \partial \vec k}\right)^{\!\!2}
      - \cB \, .
\end{align}
The spherical symmetry is required because general
principles forbid that a relativistic QFT produces 
an effective equation for a hadron mass squared outside 
a multiplet representing rotational symmetry in the 
Minkowski space. Accordingly, the $x$, $y$, and $z$ 
components of $\v k$ in Eq.~(\ref{eq:veck}) combine 
to a three-dimensional quantity that is capable of 
supporting a representation of the rotational symmetry 
in a generally valid eigenvalue formula for hadron 
masses in a relativistic theory. This happens 
irrespective of the fact that the FF of dynamics is 
developed using a specific choice of the lightlike 
vector that defines the front $x^+=0$.

It will become clear below that the parameter $\varkappa$ 
in $U_\text{eff}$ must be the same as the $\varkappa$ in 
Eq.~(\ref{eq:averaging}a). The unknown constant $\cB$ 
represents the expectation value of interactions among 
constituents within the core. 

The quadratic Ehrenfest potential appears to be in
harmony with the expectation that the potential
energy as an ingredient of the hadron mass $M$ in
the IF of dynamics increases linearly with a
distance between static quarks in the absence of
pair creation. Since the Hamiltonian eigenvalue in
the FF of dynamics is $M^2$, instead of $M$ of the
IF of dynamics, the FF potential should be
quadratic if the IF potential is
linear~\cite{Glazek:2011vg}. Since the string
picture that leads to Regge trajectories and is
supported by results of lattice simulations is
meant to be valid for large masses, our argument
for quadratic $U_\text{eff}$ can be considered
applicable also to highly excited hadron states
for which the dual soft-wall model with a quadratic
potential could be relevant. A much less commonly 
known argument for a quadratic potential is based 
on the role of conformal symmetry in introducing a 
scale in a quantum theory~\cite{deAlfaro:1976je}, 
especially in the context of LF
holography~\cite{Brodsky:2013npa,Brodsky:2013ar}. 
The theoretical argument suggests through holography 
that on the gravity side of duality the soft-wall 
model with quadratic potential deserves attention 
even if it is not perfect phenomenologically and 
requires corrections in QCD.

Our observation of correspondence between QFT 
and the Ehrenfest equation that resembles quantum 
mechanics of a single particle in a quadratic 
potential, appears to provide a direct link 
between the two sides of AdS/QFT duality. The 
link is the claim that these alternative versions 
of the theory describe the same motion around a 
multidimensional potential energy minimum using 
different variables.

%%%%%%%%%%%%%%%%%%%%%%%%%%%%%%%%%%%%%%%%%%%%%%%%%
\section{The Ehrenfest equation for form factors}
\label{sec:hff}
%%%%%%%%%%%%%%%%%%%%%%%%%%%%%%%%%%%%%%%%%%%%%%%%%

In this section we calculate a form factor of a 
hadron, $F(q^2)$, where $q$ denotes the four-momentum
transferred to a hadron by an external electroweak or
gravitational probe. Our result suggests that the 
Brodsky-deT{\'e}ramond holographic density corresponds 
to the modulus squared of our Ehrenfest function. 

We consider the form factor defined in terms of a 
matrix element of the current $\hat J^+(x=0)$ with
$q^+ = 0$. Namely,
\begin{align}
   & \bra{\text{Hadron}\!:\!P^+\!,P^\perp\!+\! q^\perp}\hat J^+(0) \ket{\text{Hadron}\!:\!P^+\!,P^\perp }  \\
   & \quad  = Q_\text{Hadron}\ 2P^+\ F(q^2) \nn \,,
\end{align}
where $Q_\text{Hadron}$ denotes the relevant charge 
and $F(0) = 1$. The current $\hat J^+$ is expressed in 
terms of the fields that correspond to the same 
scale $\lambda$ with which the Fock-space decomposition
is constructed. We arbitrarily 
simplify the theory by assuming that the effective 
constituents can be considered pointlike in the 
entire experimentally accessible range of momentum 
transfers to a hadron in elastic scattering processes
and thus we do not introduce any significant constituent 
form factors (we set them to 1). The FF form factor 
formula is~\cite{Drell:1969km,West:1970av}
\begin{align}
   & F(q^2)
   = \sum_{n}
     \int\![{\boldsymbol p}^\perp\!,{\boldsymbol x}] \sum_{j=1}^n e_j  \\
   & \quad \times 
       2(2\pi)^3\delta\Big(\sum_{i=1}^n x_i -1\Big) 
       \delta^{(2)}\Big(\sum_{i=1}^n p_i^\perp - P^\perp\Big) \nn \\
   & \quad \times
       \psi^{\dagger(n)}({\boldsymbol p_j^\perp} - {\boldsymbol x}P'^\perp\!,{\boldsymbol x};\lambda)
       \,\psi^{(n)}({\boldsymbol p^\perp} - {\boldsymbol x}P^\perp\!,{\boldsymbol x};\lambda)\,, \nn
\end{align}
where $e_j$ is the ratio of charge carried by the 
active constituent to $Q_\text{Hadron}$, the latter
assumed not zero. The momenta of $n$ constituents 
in the outgoing hadron of momentum $P'^\perp = P^\perp 
+ q^\perp$ are denoted by ${\boldsymbol p_j^\perp} = (p^\perp_i 
+ \delta_{ij}\,q^\perp)_{i=1,...,n}$. For simplicity,
we assume that $\psi^{(n)}$ is a symmetric function 
of the constituents' momenta. Then, one can choose 
a value of $j$, for instance $n$, to carry out the 
summation over active constituents. The resulting
factor $\sum_j e_j = 1$ in each and every Fock 
sector contribution is not indicated below. So,
\begin{align}
   & F(q^2) = \sum_{n} \int\![{\boldsymbol \kappa}^\perp\!,{\boldsymbol \chi}]
       \,2(2\pi)^3\delta\Big(\sum_{j=1}^{n-1} \chi_j -1\Big)
       \delta^{(2)}\Big(\sum_{j=1}^{n-1} \kappa_j^\perp \Big) \nn\\ 
   & \quad \times \int_{k,x} 
       \psi^{\dagger(n)}_{k',x} \big(\boldsymbol\kappa^\perp\!,{\boldsymbol \chi};\lambda\big)
       \,
       \psi^{(n)}_{k,x} \big(\boldsymbol\kappa^\perp\!,{\boldsymbol \chi};\lambda\big)
   \,,
\end{align}
where $k'^\perp = k^\perp + (1-x)\,q^\perp$
and $\boldsymbol\kappa^\perp = (k_j^\perp + \chi_jk^\perp)_{j=1,\dots,n-1}$
are the relative momenta of spectators  
treated as constituents of the core in 
their own center-of-mass frame.

Using the Fourier transforms of the Fock-space
wave functions $\psi^{(n)}_{k,x} \big(\boldsymbol\kappa^\perp\!,{\boldsymbol
  \chi};\lambda\big)$ in the variable $k^\perp$,
i.e., in the variable that stands in the 
subscripts, and employing the notation 
\begin{align}
\int_{\eta,x}
& =
\int\!{d^2\eta^\perp \, dx \over 4\pi\, x(1-x)} \, ,
\end{align}
one obtains the formula
\begin{align}
\label{eq:FormFactorBT}
   F(q^2)
   & = \sum_{n}
       \int\![{\boldsymbol \kappa}^\perp\!,{\boldsymbol \chi}]
       \,2(2\pi)^3\delta\Big(\sum_{j=1}^{n-1} \chi_j -1\Big)
       \delta^{(2)}\Big(\sum_{j=1}^{n-1} \kappa_j^\perp \Big)\nn\\
   & \quad \times \int_{\eta,x}
       \,e^{i (1-x) \eta^\perp q^\perp }
       \,\left|\tilde\psi^{(n)}_{\eta,x}\big(\boldsymbol\kappa^\perp\!,{\boldsymbol \chi};\lambda\big)\right|^2
   \,,
\end{align}
where $\big|\tilde\psi^{(n)}_{\eta,x}\big|^2$
contributes to the Brodsky-deT{\'e}ramond probability 
density. In terms of our Ehrenfest function,
the form factor of Eq.~(\ref{eq:FormFactorBT}) 
reads
\begin{align}
\label{eq:FormFactorE}
   F(q^2)
   & = \E{ \int_{\eta,x} e^{i (1-x)\eta^\perp q^\perp}\,
          \left|\tilde\psi^{(n)}_{\eta,x}\big(\boldsymbol\kappa^\perp\!,{\boldsymbol \chi};\lambda\big)\right|^2
       } \\
   & = \int_{\eta,x}
       \,e^{i (1-x) \eta^\perp q^\perp }
       \left|\tilde\psi(\eta^\perp\!,x)\right|^2 \, , \nn
\end{align} 
where the Fourier transform of 
the Ehrenfest function $\psi(k^\perp\!,x)$
denoted by $\tilde\psi(\eta^\perp\!,x)$ is
\begin{align}
\label{psiFT}
\tilde\psi(\eta^\perp\!,x)
   & = 
   \int { d^2k \over (2\pi)^2} \, 
e^{- i \eta^\perp k^\perp} \, \psi(k^\perp\!,x)\, .
\end{align}
In the Breit frame, we have $q^-$ $=$ $q^+$ $=0$
and the product of relevant four-vectors $q
\eta$ is reduced to $- q^\perp \eta^\perp$.
Therefore, Eq.~(\ref{psiFT}) may in principle
lead to a complete result of a fully relativistic 
theory despite that it only involves the Fourier 
transform in the transverse distance between the 
effective constituents. We see in Eq.~(\ref{eq:FormFactorE})
that the same charge distribution that is {\it a
priori} carried by the infinitely many scale-dependent
constituents in all Fock-space
components of a hadron in QFT, is equally well
representable by the modulus squared of a single
Ehrenfest function of a simple variable.

Note that we do not require that all constituents 
are charged. For example, some of them, such as 
gluons in hadrons, may have zero electric charge 
and do not contribute to a form factor (all 
constituents are meant to contribute to a 
gravitational form factor). Nevertheless, if 
the wave functions are symmetric, all constituents 
behave in the same way with respect to the 
corresponding cores and all cores have the same 
properties. Deviations from the symmetry of quantum 
wave functions imply deviations from the equivalence 
of the averaging in the form factor formula and in 
the Hamiltonian expectation value. Such deviations
ought to be accounted for in the LF holography.

%%%%%%%%%%%%%%%%%%%%%%%%%%%%%%%%%%%%%%%%%%%%
\section{ Example of the Ehrenfest picture }
\label{sec:model}
%%%%%%%%%%%%%%%%%%%%%%%%%%%%%%%%%%%%%%%%%%%%

We assume that each constituent of a hadron in 
every Fock component can be looked at as attracted 
by some force to the spectator constituents in the 
same Fock component. The interactions that mix 
different sectors, i.e., change the number of 
effective constituents of scale $\lambda$, are 
assumed fully translated to the interactions within 
the same sector as outlined in Sec.~\ref{sec:eh}. 
The translation is carried out using the condition 
that the collection of effective Fock components 
that represents a hadron satisfies the eigenvalue 
equation for the FF Hamiltonian $\hat P^-$ obtained 
at some value of $\lambda$ by integrating the RGPEP 
equations from $\lambda = \infty$ (assuming one has
found the required counterterms, also using the RGPEP). 
We now seek the picture that results from averaging 
the dynamics over all effective constituents and all 
Fock components in the effective-particle basis 
corresponding to some value of $\lambda$. 

Following the assumption that in every Fock component 
every active constituent can be seen as attracted to 
its spectators by some $\lambda$-dependent effective 
potential that reflects the dynamics of formation of 
the component in the underlying theory, we expect that
every constituent can be seen as attracted to a minimum 
of the potential energy in its component. Around its 
minimum, the effective potential in every component is 
assumed a quadratic function of the distance between 
the active constituent and the center of mass of 
spectators.

For the purpose of building this intuitive picture, 
our example may be interpreted as resembling the Thomson 
model of an atom where an electron is attracted to 
the center of a distribution of a positive charge. 
Another intuitive analogy is provided by the model 
of a nucleus where a single nucleon moves in a 
potential created by other nucleons. These old  
intuitive pictures must now be extended by a new 
element which is introduced by averaging over the 
presumably infinitely large collection of effective 
Fock components, which are defined using the RGPEP 
and which only together satisfy the FF Hamiltonian 
eigenvalue equation for the whole hadron.

%%%%%%%%%%%%%%%%%%%%%%%%%%%%%%%%%%%%%%%%%%%%
\subsection{ Transverse position variables }
\label{sec:Space}
%%%%%%%%%%%%%%%%%%%%%%%%%%%%%%%%%%%%%%%%%%%%

Let us focus first on the transverse distribution 
of constituents because the transverse arguments 
of the Ehrenfest function play the key role in 
the form factor formula for $q^+=0$. Let the 
absolute transverse position of the $i$th constituent 
be denoted by $r_i^\perp$. Consequently, the same
transverse position of a hadron as a whole in
every sector with any number of constituents 
denoted by $n$ is described by 
(cf. Ref.~\cite{Soper:1976jc})
\begin{align}
   R^\perp
   & = \sum_{i=1}^n x_i r_i^\perp \,,
\end{align}
while the position of \emph{the center of mass of 
the spectators} accompanying the constituent $j$ is 
defined by
\begin{align}
   R_j^\perp
   & = {\sum_{i\neq j} x_i r_i^\perp \over \sum_{i\neq j} x_i}  \,.
\end{align}
We denote by $\eta_i^\perp$ the relative distance
between the $i$th constituent and the hadron 
position,
\begin{align}
\label{eq:etai}
   \eta_i^\perp 
   & = r_i^\perp - R^\perp \\
   & = (1-x_i) (r_i^\perp - R_i^\perp)\, .
\end{align}
The latter equality reminds us that $\eta_i^\perp$ 
is proportional to the transverse distance between 
the $i$th constituent and the mass center of 
spectators. 

%%%%%%%%%%%%%%%%%%%%%%%%%%%%%%%%%%%%%%%
\subsection{ Symmetric wave functions }
\label{sec:Symetric}
%%%%%%%%%%%%%%%%%%%%%%%%%%%%%%%%%%%%%%%

Details of the calculations that follow 
require a specific model of the basis 
states in the Fock space and the corresponding
hadron wave functions, both of them defined at 
some scale $\lambda$. The mathematically 
simplest model we can imagine is built
assuming that all constituents are electrically 
charged scalar quanta that are bound by strong 
interactions irrespective of their charges. 
This means that all the constituents appear 
identical from the point of view of the strong
dynamics that describes their binding and we
can neglect differences in charges of the 
constituents when building their Fock-space 
wave functions. Thus, we assume that the basis 
states $\ket{n :{\boldsymbol p^\perp}\!,
{\boldsymbol x} P^+;\lambda}$ in Eq.~(\ref{eq:hadron})  
are of the form 
\begin{align}
\ket{n :{\boldsymbol p^\perp}\!,{\boldsymbol x} P^+;\lambda}
& = {1 \over \sqrt{n!}}\prod_{i=1}^n a_{p_i}^\dagger(\lambda) 
\ket{0} 
\end{align}
and the wave functions 
$\psi^{(n)}_{P}({\boldsymbol p^\perp}\!,{\boldsymbol x};\lambda)$
are fully symmetric functions of their arguments under
permutations of numbers $i = 1, \dots, n$. 

Certainly, the theory of such charged scalar 
quanta does not appear useful as far as explaining 
true hadron observables is concerned. However, 
it is useful as an example that helps in seeking 
the rules of approximating theories that do 
include the necessary other types of quanta. 
It is only the latter type of utility that we 
try to exploit by building the example discussed 
below. Thus, one can say that the goal of the 
example is only to illustrate the idea of Ehrenfest 
correspondence between QFT and its holographic 
approximation in terms of a plausible sketch, 
making it as simple as possible but including 
the difficulty of dealing with a renormalized 
theory in the FF Fock space. 

Let the hadron ground-state wave function
in the Fock component of $n$ effective 
constituents correspond to motion around 
a potential energy minimum. We identify 
the concept of such minimum using the 
transverse position of a constituent with 
respect to the center of mass of a hadron
$\eta_i^\perp$ defined in Eq.~(\ref{eq:etai}). 
The issue is then precisely what function 
of the variable $\eta_i^\perp$ we should use 
for modeling the motion of constituents 
around the relevant minimum. The ambiguity 
concerns only the coefficient of $\eta_i^{\perp \, 2}$ 
in the exponent of a suitable Gaussian wave 
function. It will be demonstrated below that 
the choice corresponding to the Ehrenfest 
function that matches the Brodsky-deT{\'e}ramond 
holography has the form
\begin{align}
   \label{eq:model_state_position}
   &\tilde \psi^{(n)}({\boldsymbol \eta}^\perp\!,{\boldsymbol x};\lambda) 
    = \varkappa_n^{\,n}\ \tilde A_n(\lambda) \\
   & \hspace{3em} \times \exp \left\{
         - {1\over 2}\sum_{i=1}^n\left[
            x_i\,\varkappa_n^2\, \eta_i^{\perp\, 2}
            + {m_n^2 \over x_i \varkappa_n^2}
       \right]\right\} \!,\nn 
\end{align}
where $\tilde A_n(\lambda)$ is related
to the probability amplitude for finding in a hadron
the $n$ constituents corresponding to the scale 
$\lambda$. The parameter $\varkappa_n = \varkappa_n
(\lambda)$ determines the width of the $n$-particle 
wave function in the variables $\sqrt{x_i} \, 
\eta^\perp_i, i = 1, \dots, n$. Our demonstration 
below will provide an explanation of the factor
$\sqrt{x_i}$ on the basis of the argument that 
the motion of constituents corresponds to the 
motion around the minimum of a suitably defined 
potential. 

The Fourier-transformed wave function of momentum 
variables has the form shown in Eq.~(\ref{eq:psin}) 
and contains the factor $\psi^{(n)}({\boldsymbol k}^\perp\!,
{\boldsymbol x};\lambda)$ given by the formula
\begin{align}
   \label{eq:model_state_momenta}
   & \psi^{(n)}({\boldsymbol k}^\perp\!,{\boldsymbol x};\lambda)
     = {A_n(\lambda) \over \varkappa_n^{\,n}} \\
   & \hspace{3em} \times
     \exp\left\{
       - \, {1 \over 2 \, \varkappa_n^2} \ \sum_{i=1}^n
       \left[
           {k_i^{\perp\,2} + m_n^2\over x_i}
       \right]
     \right\} \nn \,,
\end{align}
where the relative momenta ${\boldsymbol k}^\perp 
= (k_i^\perp)_{i=1,\dots,n}$ are defined below 
Eq.~(\ref{eq:psin}) and satisfy constraints 
in Eq.~(\ref{eq:domain}). The coefficient 
$A_n$ is related to $\tilde A_n$ by the 
Fourier transform.

%%%%%%%%%%%%%%%%%%%%%%%%%%%%%%%%%
\subsection{ Expectation values }
\label{sec:ExpModel}
%%%%%%%%%%%%%%%%%%%%%%%%%%%%%%%%%

In this section we define our averaging procedure 
that has the properties described by 
Eqs.~(\ref{eq:averaging}a)-(\ref{eq:averaging}c) and~(\ref{eq:Ueff}). 
The normalization condition of Eq.~(\ref{eq:normalization}) 
for a hadron state given in Eq.~(\ref{eq:hadron}) 
in terms of wave functions defined in 
Eq.~(\ref{eq:model_state_momenta}), yields
\begin{align}
\label{eq:normm}
   1
   & = \sum_n \int\![\boldsymbol\kappa^\perp\!,\boldsymbol \chi]
       \,2(2\pi)^3\delta\Big(\sum_{j=1}^{n-1} \chi_j - 1\Big)
     \delta^{(2)}\Big(\sum_{j=1}^{n-1} \kappa_j^\perp\Big) \\
   & \times \int_{k,x} {|A_n|^2 \over \varkappa_n^{\,2n}}
      \exp\left\{-\!\left[
          {k^{\perp\,2}\over x(1-x)}
          + {m_n^2 \over x} 
          + {\cM_{n-1}^2\over 1-x}
       \right] \Big/ \varkappa_n^2 \right\},\nn
\end{align}
where $\cM_{n-1}^2$ is given by Eq.~(\ref{eq:m_n-1}).

In normalization Eq.~(\ref{eq:normm}), the entire
internal structure of the core manifests itself
only through the mass squared, which varies from
$[(n-1)\,m_n]^2$ to infinity. This great
simplification is a property of our symmetric
Gaussian example. However, we recall that the
motion of any system around a quadratic minimum of
potential energy is described by some Gaussian and
the lowest excited states are also generally
expected to resemble some harmonic oscillator
pattern. Therefore, even if we do not know the 
Gaussian widths $\varkappa_n$ and amplitudes $A_n$ 
for all relevant values of $n$ at any given $\lambda$, 
we can still trace the consequences of averaging 
various quantities in our example and derive the 
Ehrenfest formulas that are likely to be valid for 
hadrons in QFT irrespective of the errors introduced 
by our minimal, Gaussian approximation. 

Knowing that the spectator dependence in 
Eq.~(\ref{eq:normm}) is reduced to dependence on
$\cM_{n-1}$, we can replace the integration over all 
variables $\boldsymbol\kappa^\perp$ and $\boldsymbol\chi$ 
for spectators in every component by an integral 
over a single variable $\cM^2$ with 
the phase space density~$\rho_n$ defined by
\begin{align}
   \rho_n\left(\cM^2\right )
   & = \int\![\boldsymbol\kappa^\perp\!,\boldsymbol \chi]
       \,\delta
       \Big[
          {\cM^2 }
          - {\cM_{n-1}^2(\boldsymbol\kappa^\perp\!,\boldsymbol \chi)}
       \Big]\\
   &  \quad\times
       2(2\pi)^3\delta\Big(\sum_{j=1}^{n-1} \chi_j - 1\Big)
       \delta^{(2)}\Big(\sum_{j=1}^{n-1} \kappa_j^\perp\Big)\, .  \nn
\end{align}
The hadron expectation value of any one-particle
quantity $X_n$, i.e., any quantity $X_n$ that in all 
the Fock components depends only on $k^\perp$, $x$, 
and $\cM_{n-1}$, is given by the formula
\begin{align}
\label{eq:EX}
& \E{\int_{k,x} \psi^{\dagger (n)}_{k,x}\ X_n \ \psi^{(n)}_{k,x}}\\
   & = \quad \sum_n \int\!d\cM^2 \ \rho_n(\cM^2) 
     \,\int_{k,x} X_n(k^\perp\!, x, \cM) \nn\\
   & \quad \times
   {|A_n|^2 \over \varkappa_n^{\,2n}}
      \exp\left\{-\left[
          {k^{\perp\,2}\over x(1-x)}
          + {m_n^2 \over x} 
          + {\cM^2\over 1-x}
       \right]\Big/ \varkappa_n^2 \right\}\, . \nn
\end{align}

%%%%%%%%%%%%%%%%%%%%%%%%%%%%%%%%%%%%%
\subsection{ The Ehrenfest function }
\label{sec:ModelEf}
%%%%%%%%%%%%%%%%%%%%%%%%%%%%%%%%%%%%%

Using Eq.~(\ref{eq:EX}) as a definition for the
expectation values in Eqs.~(\ref{eq:averaging}a)-(\ref{eq:averaging}c)
and (\ref{eq:Ueff}), one arrives at
Eq.~(\ref{eq:averaging_free}). Using the concept
of motion around a minimum, one considers the
Ehrenfest potential given in
Eq.~(\ref{eq:harmonic_potencial}). Subsequently,
variation of the hadron expectation value of a FF
Hamiltonian for any RGPEP parameter $\lambda$,
yields the Ehrenfest Eq.~(\ref{eq:Ehrenfest}) with
the oscillator potential. Its ground-state
solution is
\begin{align}
\label{eq:oscillatorgs}
   \psi(k^\perp\!,x)
   & =   \cN \,\,
         e^{-\,\v k^2 /(2 \, \varkappa^2 ) } \, , 
\end{align}
where the variable $\v k$ is defined in Eq.~(\ref{eq:veck})
and the constant $\cN$ denotes the normalization factor 
that can be calculated, for example, from the 
normalization of a hadron charge distribution.

An alternative appearing fully relativistic  
form of the same Ehrenfest function derived
using Eq.~(\ref{eq:Jacobi}) is
\begin{align}
\label{eq:psis}
\psi(\v k)
   & = \cN \, 
       e^{
            -(s_k - s_0)/(2 \, \varkappa^2)
          } \, ,
\end{align}
where the $s$-channel Mandelstam invariants 
are defined by
\begin{align}
s_0   & = (m_\text{active} + m_\text{core})^2 \, , \\
s_k   & = \left(p_\text{active} + p_\text{core}\right)^2 \\
      & =
{k^{\perp\,2} + m_\text{active}^2 \over x} 
+
{k^{\perp\,2} + m_\text{core}^2\over 1-x} \, .
\end{align}
This result means that the Ehrenfest active
constituent and core can be considered 
on-mass-shell particles with masses 
$m_\text{active}$ and $m_\text{core}$, 
while the Ehrenfest function depends only 
on their total invariant mass squared. 

One should remember at this point that the 
FF of Hamiltonian dynamics requires that the 
difference of the four-momenta of a hadron 
with mass squared eigenvalue $M^2$ and 
its Ehrenfest constituents is lightlike,
\begin{align}
\label{eq:Pn}
p_\text{active}^\mu + p_\text{core}^\mu 
& = P^\mu + { s_k - M^2 \over P^+ } \, n^\mu \, ,
\end{align}
where the null-vector $n$ defines the
front $x^+=0$ by the condition $nx=0$
and its only nonzero component is $n^-=2$.
On the other hand, Eq.~(\ref{eq:Pn}) shows 
that in the IMF, where the parton model is 
defined in the limit $P^+ \rightarrow 
\infty$, the Ehrenfest active constituent 
has the interpretation of the Feynman 
parton~\cite{Feynman:1969ej}. The active 
constituent carries the fraction $x$ and 
the core carries the fraction $1-x$ of the 
hadron momentum. The limit $P^+ \rightarrow 
\infty$ for the Ehrenfest constituents exists 
because the averaged effective dynamics does 
not allow $s_k$ to take large values, see 
Eqs.~(\ref{eq:oscillatorgs}) and (\ref{eq:psis}). 
In particular, the dominant part of the transverse 
motion of the Ehrenfest partons is soft, since it 
is limited by the scale $\varkappa$ associated with 
$\Lambda_{QFT}$. Only the large-$k^\perp$ tail
of the transverse distribution may be sensitive 
to the hard processes driven by the underlying
local theory.

The Ehrenfest function interpretation discussed 
above closely resembles the two-body quantum 
wave function interpretation of LF holography in 
the case $m_\text{active} = m_\text{core} = 
0$~\cite{Brodsky:2007hb}. However, the 
phenomenologically useful holographic wave 
functions, for example, see Refs.~\cite{Ahmady:2012dy,
Chabysheva:2012fe,Forshaw:2012im}, often operate
with a generalization to constituents with 
nonzero masses. The Ehrenfest function
provides a natural explanation of the presence 
of masses in the parton densities in the
forms advocated on phenomenological grounds.
The masses enter through the dynamics in the
$z$ direction, which complements the transverse
dynamics in our model in agreement with the 
requirement of rotational symmetry as a part of 
the Poincar\'e symmetry in QFTs and their duals.

%%%%%%%%%%%%%%%%%%%%%%
\section{ Conclusion }
\label{sec:conclusion}
%%%%%%%%%%%%%%%%%%%%%%

When the renormalized Schr\"odinger Hamiltonian 
eigenvalue equation for a ``hadron'' state in 
QFT in the FF of dynamics is reduced to its 
expectation value in any of the eigenstates 
corresponding to smallest masses for a fixed 
set of other quantum numbers, one obtains the 
Ehrenfest-like equation that describes a 
semiclassical function $\psi(\v k)$, or its 
Fourier transform $\tilde \psi(\v \eta)$ that
describes the charge density in a hadron
according to the formula 
\begin{align}
\rho%_\text{Hadron}
(\v \eta) & = Q_\text{Hadron} \, |\tilde \psi(\v \eta)|^2 \, .
\end{align}
This interpretation follows from the Ehrenfest formula
for hadron form factors. 

The Ehrenfest form factor formula matches the 
Brodsky-deT{\'e}ramond LF holography formula for the 
form factors. Thus, the principle of correspondence 
between quantum and classical dynamics discovered by 
Ehrenfest~\cite{Ehrenfest}, appears to provide a 
link between the quantum field theoretic Fock-space 
picture of ground states or slightly excited states 
of hadrons and the semiclassical gravitational picture 
suggested for them by the AdS/QFT duality~\cite{Maldacena:1997re,
Polchinski:2002jw,Polchinski:2001tt}. While the duality 
based on some M/string theory requires investigation, 
the Ehrenfest correspondence between quantum and classical 
theories is already well established in other areas 
of physics in the low-energy domain and should be 
falsifiable in particle physics in the high-energy 
domain.

If the Ehrenfest function is expected to capture 
in its shape the parton distributions for all choices 
of the momentum transfer from an external probe to
an active constituent, the function must fall off at 
large momenta in a way that is sensitive to the Fock 
components with constituents of large virtuality, 
while for small $\v k$ it may still have the shape
resembling Gaussian. In particular, for small momenta 
$\v k$ one may expect resemblance of $\psi(\v k)$ 
to the constituent quark model wave functions 
(quark-antiquark for mesons and quark-diquark in 
baryons), while for the large $\v k$ one must 
expect the corrections to a Gaussian shape that 
reflect the influence of high-energy FF Fock-space 
wave functions on the Ehrenfest expectation value. 
The high-energy behavior of wave functions can be 
studied using perturbative methods~\cite{Lepage:1980fj}. 

It should also be remembered, as mentioned in 
Sec.~\ref{sec:Averging}, that the observables such 
as elastic form factors result from coupling of the 
external probes to the effective constituents in a 
way that depends on the momentum scale of external 
probes. Only in the approximation that the current 
operators of effective constituents are independent 
of the momentum transfer from external probes, as we 
assumed for simplification in this paper, one can 
ignore the corrections that are not of the Gaussian 
type. This also means that one should expect
corrections to the soft-wall models with quadratic
potential in the bulk on the gravity side of
duality concerning QCD.

The Ehrenfest description of hadrons in Gaussian
approximation and LF holography may be explicitly
related to each other using the resemblance
between the three-dimensional Ehrenfest equation
with harmonic potential and the
Brodsky-deT{\'e}ramond two-dimensional eigenvalue
equation with a transverse harmonic oscillator
potential associated with a corresponding AdS
dilaton warping. Separating variables in
Eq.~(\ref{eq:Ehrenfest}), one obtains solutions in
the form 
\begin{align}
\label{eq:psivk}
   \psi(\v k)
   & = \phi(k_x,k_y)\ H_{n_z}\left({k_z \over \varkappa}\right)
       e^{-k_z^2/2\varkappa^2}\,,
\end{align}
where $H_{n_z}$ denotes a Hermite polynomial.
Assuming that the transverse motion corresponds 
to the angular momentum projection on the $z$ axis
equal $l_z$, one arrives at
\begin{align}
\label{eq:AdS}
   & \hspace{-1em} \Bigg[
          \!-\!\left({\partial \over \partial \zeta}\right)^2\!
          - {1\over \zeta}{\partial \over \partial \zeta}
          + {1 \over \zeta^2}l_z^2 
 %  & \quad \quad 
          + \left(2n_z + 1\right)\varkappa^2
          + \varkappa^4 \zeta^2   \\
   & \quad\, 
     - \cB +  (m_\text{active} + m_\text{core})^2
       \Bigg]
       \tilde\phi(\zeta)
     =  M^2 \ \tilde\phi(\zeta) \nn\, . 
\end{align}
The function $\tilde \phi(\zeta)$ is the radial
factor in the Fourier transform of $\phi(k_x,k_y)$
and $\zeta = |\zeta^\perp|$ is the length of the 
Brodsky-deT{\'e}ramond holography transverse position 
variable understood as 
\begin{align}
\zeta^\perp 
& = 
\sqrt{x(1-x)} \, 
(r^\perp_\text{active} - r^\perp_\text{core}) \, ,
\end{align}
where variables $r^\perp_\text{active}$ and $r^\perp_\text{core}$
denote the transverse positions of the parton and 
core on the front. Thus, $\zeta^\perp$ is the relative 
transverse position of the active parton and hadron 
core rescaled by $\sqrt{x(1-x)}$, where $x$ is the 
hadron-momentum fraction carried by the active 
parton in the IMF. 

Equation~(\ref{eq:AdS}) precisely matches the holography 
equations, e.g., Eq.~(11) in Ref.~\cite{deTeramond:2008ht} 
or Eq.~(33) in Ref.~\cite{deTeramond:2013it}, after 
rescaling $\tilde\phi(\zeta)$ by $\sqrt \zeta$ and 
adjusting the additive constant $(m_\text{active} + 
m_\text{core})^2 - \cB$. 

Unfortunately, our simple consideration does 
not uniquely predict the value of the constant $\cB$ 
for various hadrons and instead merely indicates a 
need for its existence. Since the constant $\varkappa$ 
in the Ehrenfest $U_\text{eff}$ may be thought of as related 
to the gluon condensate inside hadrons~\cite{Glazek:2011vg}, 
one may expect that an explanation of the constant 
$\cB$ in $U_\text{eff}$ also requires understanding 
of the dynamics that involves distances comparable with 
the size of a hadron.

Explanation of the Ehrenfest harmonic potential 
in a full theory must account for the interaction 
between the active constituent and core including 
a form factor of the core, say 
$f(\v q^2)$, where $\v q = \v k' - \v k$ is the 
momentum transfer between the active constituent 
and the core. The core form factor is to describe
the core structure. The validity of this picture 
is expected by analogy with approximations such 
as the Hartree-Fock approximation to many-body 
dynamics. However, in the case of QFT the situation 
is greatly complicated due to the interactions that 
are capable of violent changes of energies and 
extensive mixing of various numbers of virtual 
constituents. Nevertheless, for sufficiently big 
numbers of effective constituents, which means 
sufficiently large $\lambda$, the core is similar 
in its strong-interaction charge or inertia 
distribution to the hadron itself. So, in the 
first approximation,
\begin{align}
   f(\v q^2) \sim F(q^2) \, ,
\end{align}
where $F(q^2)$ is the measurable hadron form factor 
and the proportionality refers to the proper charge 
coefficient. Although $F(q^2)$ does not describe 
the distribution of neutral constituents, such as 
gluons in the case of electromagnetic form factors 
of a proton, while $f(q^2)$ describes the density 
of all strongly interacting constituents that form 
the core, which includes the gluons, the average 
distributions are likely to be of essentially similar 
shape. Establishing that they are not would constitute 
a new insight into the structure of hadrons.

Assuming a considerable range of validity of the 
Gaussian approximation, the main limitation in the 
accuracy of representing the QFT theory solutions 
with the Ehrenfest function stems from the fact 
that the averaging involved in evaluating expectation 
values misses the interference effects that can be 
fully described only using the Fock-space wave 
functions. On the other hand, it can be checked 
to what extent the full Ehrenfest function may 
describe the parton distributions in more detail 
than may its modulus squared alone. If it did, it would
offer a semiclassical single-parton picture that 
would stand a chance of approximating the physics 
of hadrons a bit more accurately than the entirely 
probabilistic description of the parton-model
type, or of the type of quantum mechanical models 
that are not directly related to QFT. 

We hasten to conclude that the duality of an AdS-like 
picture to QFT can be interpreted in terms of the 
Ehrenfest function. The behavior of the 
Ehrenfest potential could thus 
correspond to the behavior of potentials in 
duality models such as in Ref.~\cite{Karch:2006pv}.
Needless to say, similar considerations should apply 
in principle to the discussion of many QFTs.

\begin{acknowledgments}
The authors thank Stan Brodsky for recent
discussions. This work was supported by the 
Foundation for Polish Science International 
Ph.D. Projects Programme cofinanced by the 
EU European Regional Development Fund.
\end{acknowledgments}

\bibliography{bibliography}
\end{document}